# Superconductivity-induced Magnetic Modulation in a Ferromagnet Through an Insulator in $La_{2/3}Ca_{1/3}MnO_3$/$SrTiO_3$/$YBa_2Cu_3O_{7-\delta}$ Hybrid Heterostructures

C. L. Prajapat[1], Surendra Singh[2], Amitesh Paul[3], D. Bhattacharya[2], M. R. Singh[1],

G. Ravikumar[1] and S. Basu[2,*]

[1]*Technical Physics Division, Bhabha Atomic Research Centre, Mumbai-400085.*

[2]*Solid State Physics Division, Bhabha Atomic Research Centre, Mumbai-400085.*

[3]*Technische Universität München, Physik Department, Lehrstuhl für Neutronenstreuung, James-Franck-Straße 1,D-85748 Garching, Germany*

*email: sbasu@barc.gov.in

Coexistence of ferromagnetic and superconducting orders and their interplay in ferromagnet-superconductor heterostructures[1-5] is a topic of intense research. While it is well known that proximity of a ferromagnet suppresses superconducting order in the superconductor, there exist few studies indicating the proximity of a superconductor suppressing ferromagnetic order in a ferromagnet.[6-11] Here we demonstrate a rare observation of the suppression of ferromagnetic order in a $La_{2/3}Ca_{1/3}MnO_3$ layer separated from a $YBa_2Cu_3O_{7-\delta}$ layer by a thin insulator ($SrTiO_3$). Polarized neutron reflectivity measurements on $La_{2/3}Ca_{1/3}MnO_3$/$SrTiO_3$/$YBa_2Cu_3O_{7-\delta}$ trilayer deposited on [001] $SrTiO_3$ single crystal substrates shows the emergence of a thin magnetic "dead" layer in $La_{2/3}Ca_{1/3}MnO_3$ adjacent to the insulating layer below its superconducting transition temperature of $YBa_2Cu_3O_{7-\delta}$. Further, the magnetic dead layer grows in thickness when the **insulating layer is made thinner. This indicates a possible tunneling of the superconducting order-parameter through the insulating SrTiO$_3$ inducing modulation of magnetization in La$_{2/3}$Ca$_{1/3}$MnO$_3$.**

Superconductors (SC) and ferromagnetic (FM) heterostructure like SC/FM/SC, are known to exhibit coupling between the SC through a thin intervening FM layer producing what is known as π-state, essentially due to the presence of SC order-parameter in the FM layer, albeit over an extremely short range.[12] However, in thin film hetrostructures of SC cuprates and FM manganites, there are sufficient evidences which suggest that the SC order persists over much longer length scales extending up to 100 Å,[12,13] which is very intriguing by itself. YBa$_2$Cu$_3$O$_{7-\delta}$ (YBCO) and La$_{2/3}$Ca$_{1/3}$MnO$_3$ (LCMO) are ideal candidates for growth of epitaxial thin films on a variety of oxide substrates like SrTiO$_3$ (STO), LaAlO$_3$ and others. Heterostructures of these materials grown with high interface quality[14,15] are ideal candidates for investigating the interaction between mutually antagonistic SC and FM orderings. The properties of LCMO/YBCO heterostructures are strongly influenced by a coupling phenomena at the interface,[1-3] which can lead to complex behaviors in these heterostructures such as giant magnetoresistance[4], transient photo-induced superconductivity[5] and magnetic proximity effects.[6-11] While the competition between the two ordered ground states[1-3,16] leads to suppression of both superconducting and magnetic transition temperatures,[17,18] a variety of exotic phenomenon have been seen in these heterostrures.[19]

Hoppler *et al.*,[9] inferred a giant modulation of the in-plane magnetization in LCMO layers below the superconducting transition of LCMO/YBCO multilayer. It was observed that the magnetization in alternate LCMO layers are strongly suppressed and enhanced, doubling the periodicity of the magnetic lattice. There also exist experimental studies which indicated the depletion of magnetization or a magnetic dead (MD) layer in the adjacent region of LCMO at the LCMO/YBCO interface.[6-11] Magnetic dead layers are known to result from chemical inter-diffusion/alloying or interface roughness[20-22]. However, scattering techniques[6-11] and electron spectroscopy[23] have ruled out these factors in LCMO/YBCO heterostructures. While magnetic modulation in LCMO layers, induced by superconductivity in the adjacent YBCO layer in LCMO/YBCO multilayers have been extensively studied[5-9], to the best of our knowledge no study involving superconducting order-parameter influencing a ferromagnet through an insulating (I) barrier in FM/I/SC hetrostructures has been reported so far. Here we observed tunneling effect across an intermediate insulator layer and found that thickness of the insulating layer play an important role for deciding the behavior of such hybrid system.

Two trilayer samples labeled as S1: STO[substrate]/YBCO (300 Å) / STO (25 Å)/ LCMO (300 Å) and S2: STO[substrate]/YBCO (200 Å) / STO (50 Å)/ LCMO (200 Å) were grown with STO (100) as substrate by pulse laser deposition (PLD). Using polarized neutron reflectivity (PNR), we present direct evidence of magnetic modulation in LCMO layer across insulating STO layer below superconducting transition temperature ($T_{SC}$). PNR data reveals that the magnetization in LCMO was suppressed to zero (magnetic "dead" layer) near the LCMO/STO interface below $T_{SC}$. The thickness of the magnetic "dead" layer is estimated to be about 100 Å in sample S1. The magnetic dead layer thickness is reduced to ~ 40 Å in sample S2 where the thickness of insulator was increased to 50 Å. This clearly signifies the tunneling of the SC order-parameter through an insulator into a FM.

The structural characterization of the samples was done by using X-ray diffraction (Fig. S1 in Supplementary), showing high quality epitaxial growth of the films. X-ray reflectivity (XRR) measurements were performed to determine the depth dependent layer structure of these hetrostructures (Supplementary Fig. S2). Fig. 1 depicts SQUID data for the d. c. magnetization measurements on S1 under field cooled (FC) condition (cooling field $H_{FC}$ = 300 Oe). The zero field cooled (ZFC) SQUID data are shown in the inset of Fig.1. The ZFC data shows $T_{SC}$ ~ 60 K, suggesting the YBCO is under-doped and FC data shows the LCMO layer has a Curie temperature ~ 150 K. Similar behaviors for SQUID data from S2 was also observed.

PNR measurements were carried out to obtain depth dependent magnetization profile in the samples. PNR involves specular reflection of polarized neutron from magnetic film as a function of wave vector transfer, $Q$ (= $4\pi \sin\theta/\lambda$, where, $\theta$ is angle of incidence and $\lambda$ is neutron wavelength).[22,24,25,26] Specular reflection of neutron beam with polarization parallel (+) and anti-parallel (-) to sample magnetization corresponds to reflectivities, $R^{\pm}(Q)$. We have measured the PNR data for S1 and S2 at 10 K, 50 K, 100 K and 300 K, with an applied in-plane field of 300 Oe after cooling the sample in the same field from 300 K. Our aim was to obtain the magnetization depth profile from the fits to these data sets and look for any possible modulations in magnetization across $T_{SC}$. Fig. 2a shows the PNR measurements from S1 at 300 K and 10 K. The normalized spin asymmetry (NSA) plots at 300 K and 10 K shown in the bottom panel of Fig. 2a is given by $(R^+ - R^-)/R_F$, where $R_F = 16\pi^2/Q^4$ is Fresnel reflectivity.[24] At 300 K, $R^+$ and $R^-$ are same (NSA = 0), indicating no net magnetization of the sample at this temperature, consistent with the macroscopic magnetization (SQUID) measurements.

In order to extract the magnetization profile from the PNR measurements,[25] we first optimized the nuclear scattering length density (NSLD) profile at 300 K by constraining the layer thicknesses, density and interface roughness to be within the error estimated on parameters obtained from XRR. Keeping the NSLD fixed, the magnetization depth profile [$M(z)$] was optimized to fit the PNR data at lower temperatures. Since the effect that we observed is more enhanced for thinner intervening insulator (STO) layer, we will first discuss the results obtained from S1, followed by the results from S2. The solid lines in Fig. 2a represent the fit to experimental data at 300 K and 10 K, from S1. Top panel of Fig. 2b shows the NSLD obtained from PNR data at 300 K.

The PNR data from S1 at 10 K (Fig. 2a) shows a clear separation between $R^+$ and $R^-$ indicating a ferromagnetic state of the LCMO layer (contrast with the data at 300 K in the same figure). To fit the PNR data at 10 K, we have optimized several models for the magnetization depth profile in the LCMO layer and a detailed comparison of the fits corresponding to different models is shown in the Supplementary Fig. S3. Uniform magnetization profile (dash blue line in the bottom panel of Fig. 2b) for the entire LCMO layer clearly does not agree with the NSA data at 10 K as shown by the dash (blue) lines in Fig. 2a (bottom panel). The fit (black and green curves in top panel of Fig. 2a) to PNR data at 10 K with minimum $\chi^2$ was obtained for a magnetization depth profile shown as a solid (black) curve in the bottom panel of Fig. 2b, indicating a magnetic "dead" layer (shaded area in bottom panel of Fig. 2b) at LCMO/STO interface with a thickness $\Delta \sim 100$ Å and a non-uniform magnetization in the rest of the LCMO layer. The magnetization in the rest of the LCMO layer is seen to be gradually increasing from zero to a maximum value ~205 emu/cm$^3$ near the film-air interface.

To confirm the role of superconductivity on the modulation of the magnetization depth profile of LCMO layer in S1 we carried out PNR (NSA) measurements at 100 K (well above the $T_{SC}$) and 50 K (marginally below $T_{SC}$~60 K) as shown in top and bottom panel of Fig. 3a, respectively. We attempted to fit PNR data from S1 at 100 K with similar magnetization depth profile as obtained at 10 K. The fit is shown in Fig. 3b (top panel) as solid curve. Also a fit with uniform magnetization in the LCMO layer without any dead-layer, is shown in the top panel of Fig. 3b as blue dashed line. Comparison of these two fits clearly suggests that a small but uniform magnetization (dash line) of ~17 emu/cm$^3$ for the entire LCMO layer fits the NSA data better than the model with MD layer (solid curve). However the PNR data at 50 K (bottom panel of Fig. 3a and 3b) is consistent with the magnetization (with reduced magnetization) model obtained at 10 K and fits the PNR data with the same profile as obtained from the PNR data at 10 K. A comparison of fitted NSA data and magnetization depth profile model at 50 K is given in the bottom panel of Fig. 3a and b. Overall decrease in the magnetization at 50 K as compared to 10 K indicates the change in magnetization of a ferromagnetic material with temperature. These results clearly suggest that the MD layer emerges only below the superconducting transition temperature of YBCO (~ 60 K).

Further, to study the effect of insulator thickness on SC induced magnetization modulation in LCMO layer, we now focus on the result of the PNR experiments at 10 K under similar conditions on S2 (with STO layer thickness 50 Å). The sample was also characterized using XRR for depth dependent scattering length density profile and the details are given in the Supplementary information (Fig. S2). Fig. 4 shows the PNR (NSA) data from S2 at 300 K and 10 K. At 300 K we did not observe any difference between $R^+$ and $R^-$, indicating that the LCMO layer was non-magnetic (similar to S1 at 300 K). The PNR data at 300 K was analyzed to get a detailed NSLD profile of the sample (top panel of Fig. 4 a-b). PNR data at 10 K and a comparison of fit assuming different magnetization model is shown in Fig. S5 of supplementary information. The PNR (NSA) data at 10 K (bottom panel of Fig. 4 a-b) confirms the modulation of magnetization with occurrence of magnetic dead layer at LCMO/STO interface. We obtained a magnetic dead layer of thickness ~ 40 Å at the LCMO/STO interface of S2. The magnetization profile in the rest of the LCMO layer was uniform with $M = 100$ emu/cm$^3$.

At this point it is relevant to mention that the length scale of the MD layer at LCMO/STO interface is much higher (≈ 40-100 Å) than the interface roughness (≈ 5 Å) as obtained from the XRR and PNR data. The interface roughness in these samples ranged between 13 Å and 4 Å, the highest being at the air film interface (13 ± 3 Å) and the lowest at the LCMO on STO interface (4 ± 1 Å). This fact overrules interface mixing as a possible cause of the observed dead-layer.

In addition, the average magnetization of S1 was estimated around 120 ± 10 emu/cm$^3$ at 10 K using PNR measurements, which is comparable with earlier measurements[7] and well below the saturation magnetization ($M_s$) ≈ 400 emu/cm$^3$ observed for single layer LCMO thin films.[27] This is also in agreement with the value of 112 emu/cm$^3$, obtained from SQUID by considering a magnetic dead layer of thickness ~ 100 Å. It has been reported that magnetization in LCMO depend on the thickness of YBCO layers.[17] We believe that lower value of $T_{SC}$, Curie temperature and magnetic moment in this system is an important issue for observed results, because these may well provide an additional energy and length scale that must be considered in describing the competing SC and magnetic interactions.[28]

Our PNR results from LCMO/STO/YBCO systems clearly indicate that the existence of a MD layer is related to the superconducting state of YBCO layer. It is a distinct possibility that the depletion of magnetization in the LCMO layer is caused by the tunneling of SC order-parameter into the LCMO layer. Both FM and the SC states derive their existence from the local density of states. Possibility of long range coherence length of Cooper pairs in FM has been theoretically discussed by Bobkova and Bobkov.[29] We argue that the system prepared under field cooled condition is in a non-equilibrium state which comprises a nanoscopic phase-coexistence of FM, AFM and charge ordered states and FM domain walls. We suspect that the LCMO layer which show coexistence[30, 31,32] of different phases mentioned above will provide a spatially varying characteristic length scales for SC order-parameter and thus play a fundamental role for observation of such SC-induced modulation in magnetization. The decay of such superconducting wave functions in LCMO layer after tunneling through STO is depicted in Fig. 5a. This decay of SC order in the FM layer, in our samples, is indicated by the gradual increase of the magnetic moment density profile. In view of this we fitted (Fig. 5 b) the magnetization depth profile of S1 at 10 K obtained from PNR measurements to the expression: $M(x) = M_0[1 - \exp\left(-\frac{x}{\xi}\right)]^\alpha$, where $x$ is distance of surface (air/LCMO interface) from YBCO/STO interface, $\xi$ is coherence length scale, $M_0$ magnetization at surface and $\alpha$ is an exponential coefficient, which will dictate the possible order of length scale involved in LCMO layer. From the fitted line (Solid line in Fig. 5 b), the exponential ($\alpha$) for the system is estimated at around 18: a surprisingly large value! This indicates the presence of a large number of length scales in the system.

In conclusion, we have shown unambiguously that superconductivity induced modulation in magnetization depth profile in LCMO layer across an insulating STO layer in two LCMO/STO/YBCO hybrid structures. We observed that a magnetic dead layer formed at the LCMO/STO (LCMO on STO) interface below superconducting transition temperature of YBCO. We conjecture that this happens probably due to tunneling of superconducting order-parameter in to the FM layer. The length scale for the magnetic dead layer depends on the thickness of the insulator layer. Thinner the insulator layer, thicker was the magnetic dead layer at LCMO/STO interface. We believe the presence of phase coexistence over many length scales in LCMO layer is responsible for superconductivity-induced magnetic modulation in these hetrostructures. Our results opens a way to explore the fundamental study of tunneling of superconducting order parameter in FM/I/SC system. Nevertheless future experiments using advanced local imaging techniques in combination of scattering techniques may provide further insight for superconducting induced phenomena in FM/I/SC systems.

**Methods:**

Two trilayer samples labeled as S1: STO[substrate]/YBCO (300 Å) / STO (25 Å)/ LCMO (300 Å) and S2: STO[substrate]/YBCO (200 Å) / STO (50 Å)/ LCMO (200 Å) were grown on STO (001) substrate using pulse KrF laser (248 nm) laser deposition (PLD). During growth, the substrate temperature was 770°C, $O_2$ partial pressure was 0.5 mbar, laser fluence was 2.5 J/cm$^2$, and the pulse repetition rate was 2 Hz.

Magnetization measurements were performed using superconducting quantum interface device (SQUID) magnetometry under field cooled (FC) and zero field cooled (ZFC) conditions.

Polarized neutron reflectivity (PNR) measurements of the samples were carried out using the polarized reflectometer MARIA at the FRM II research reactor in Garching, Munich. In PNR the intensity of the specularly reflected neutron beam was measured as a function of wave vector transfer, $Q$ (= $4\pi\sin\theta/\lambda$, where, $\theta$ is angle of incidence and $\lambda$ is neutron wavelength), and for neutron beam polarization parallel (+) and anti-parallel (-) to sample magnetization. The specular reflectivity, $R$, is determined by the neutron scattering length density (SLD) depth profile, $\rho(z)$, averaged over the lateral dimensions of the sample.[22,24] $\rho(z)$ consists of nuclear and magnetic SLDs such that $\rho^{\pm}(z) = \rho_n(z) \pm CM(z)$, where $C = 2.91 \times 10^{-9}$ Å$^{-2}$ cm$^3$/emu and $M(z)$ is the magnetization (a moment density obtained in emu/cm$^3$) depth profile.[24] The +(-) sign denotes neutron beam polarization parallel (opposite) to the applied field and corresponds to reflectivities, $R^{\pm}(Q)$. Thus, by measuring $R^+(Q)$ and $R^-(Q)$, $\rho_n(z)$ and $M(z)$ can be obtained separately. Normalized spin asymmetry (NSA) is defined as $(R^+ - R^-)/R_F$, where $R_F$ is Fresnel reflectivity. It is used to enhance the role of magnetization depth profile in our analysis.

**Acknowledgements:**

We are thankful to S. Mattauch for assisting in the PNR measurements and P. Böni for his encouragements during the course of the work

**Author Contributions:**

C. L. P., M. R. S. and G. R. grew samples by pulsed laser deposition and carried out SQUID measurements. S. S., D. B. and S. B. designed the XRR and neutron scattering experiments as well as analyzed the results. A. P. proposed for the beamtime and carried out the neutron reflectivity measurements and discussed the results. All the authors contributed in writing the paper. Authors C.L.P. and S.S. have equal contributions of the work.

**Competing Interests:**

The authors declare that they have no competing financial interests.

**Figure Captions:**

Fig. 1: **Temperature dependent Magnetization data.** Magnetization data of the YBa$_2$Cu$_3$O$_{7-\delta}$ (300 Å)/SrTiO$_3$ (25 Å)/La$_{2/3}$Ca$_{1/3}$MnO$_3$ (300 Å) sample in field cooled (FC) condition in a field of ~300 Oe showing the FM transition ($T_C$) ~ 150 K. Inset show the zero field cooled (ZFC) data suggesting a superconducting transition temperature ($T_{SC}$) ~ 60 K.

Fig. 2: **Polarized neutron reflectivity (PNR) measurements and their modeling. a,** PNR (spin up, $R^+$ and spin down, $R^-$) data from the YBa$_2$Cu$_3$O$_{7-\delta}$ (300 Å)/SrTiO$_3$ (25 Å)/La$_{2/3}$Ca$_{1/3}$MnO$_3$ (300 Å) sample at 300 K and 10 K, with an applied in-plane field of 300 Oe after cooling the sample in the same field from 300 K. Reflectivity data at 300 K and 10 K are shifted by a factor of 20 for the sake of clarity. Normalized spin asymmetry (NSA) data, defined as $(R^+ - R^-)/R_F$, where $R_F = 16\pi^2/Q^4$ is Fresnel reflectivity, at 300 K and 10 K (bottom panel of **a**). **b,** Nuclear scattering length density (NSLD) and magnetization (*M*) depth profile extracted from fitting PNR data at 300 K and 10 K. Two magnetization models, with and without magnetic dead (MD) layer at LCMO/STO interface, at 10 K are also depicted in **b** (bottom panel) and the corresponding fits to PNR data are shown in **a** (bottom panel).

Fig. 3: **Polarized neutron reflectivity (PNR) measurements and their modeling across superconducting transition temperature. a,** normalized Spin asymmetry (NSA) data from the YBa$_2$Cu$_3$O$_{7-\delta}$ (300 Å)/SrTiO$_3$ (25 Å)/La$_{2/3}$Ca$_{1/3}$MnO$_3$ (300 Å) sample at 100 K (top panel) and 50 K (bottom panel), with an applied in-plane field of 300 Oe after cooling the sample in the same field from 300 K. **b,** magnetization (*M*) depth profile at 100 K (top panel) and 50 K (bottom

panel) which are fitted to NSA data shown in **a**. Comparison of two magnetization models at 100 K and 50 K are depicted in **b** and the corresponding fits to PNR data are shown in **a**.

Fig. 4: **Polarized neutron reflectivity (PNR) measurements and their modeling from S2 with insulator layer of double thickness. a**, Normalized spin asymmetry (NSA) data from the YBa$_2$Cu$_3$O$_{7-\delta}$ (220 Å)/SrTiO$_3$ (50 Å)/La$_{2/3}$Ca$_{1/3}$MnO$_3$ (190 Å) sample (S2) at 300 K (top panel) and 10 K (bottom panel), with an applied in-plane field of 300 Oe after cooling the sample in the same field from 300 K.. **b**, Nuclear scattering length density (NSLD) (top panel) and magnetization (bottom panel) depth profile extracted from fitting PNR data at 300 K and 10 K as shown in **a**. Two magnetization models at 10 K are depicted in **b** (bottom panel) and the corresponding fits to PNR data are shown in **a** (bottom panel).

Fig. 5: **Schematic of tunneling of Cooper pair across insulator**. **a**, Schematic showing representation of LCMO/STO/YBCO system with phase coexistence of different length in LCMO layer which is deciding the perturbation of superconducting wave functions tunneled through STO (insulator). **b**, Fitting of magnetization depth profile of YBa$_2$Cu$_3$O$_{7-\delta}$ (300 Å)/SrTiO$_3$ (25 Å)/La$_{2/3}$Ca$_{1/3}$MnO$_3$ (300 Å) sample at 10 K obtained from PNR data using expression: $M(x) = M_0[1 - \exp\left(-\frac{x}{\xi}\right)]^\alpha$, where $x$ is distance of surface (air/LCMO interface) from YBCO/STO interface, $\xi$ is coherence length scale, $M_0$ magnetization at surface and $\alpha$ is an exponential coefficient, which will dictate the possible order of length scale in LCMO layer.

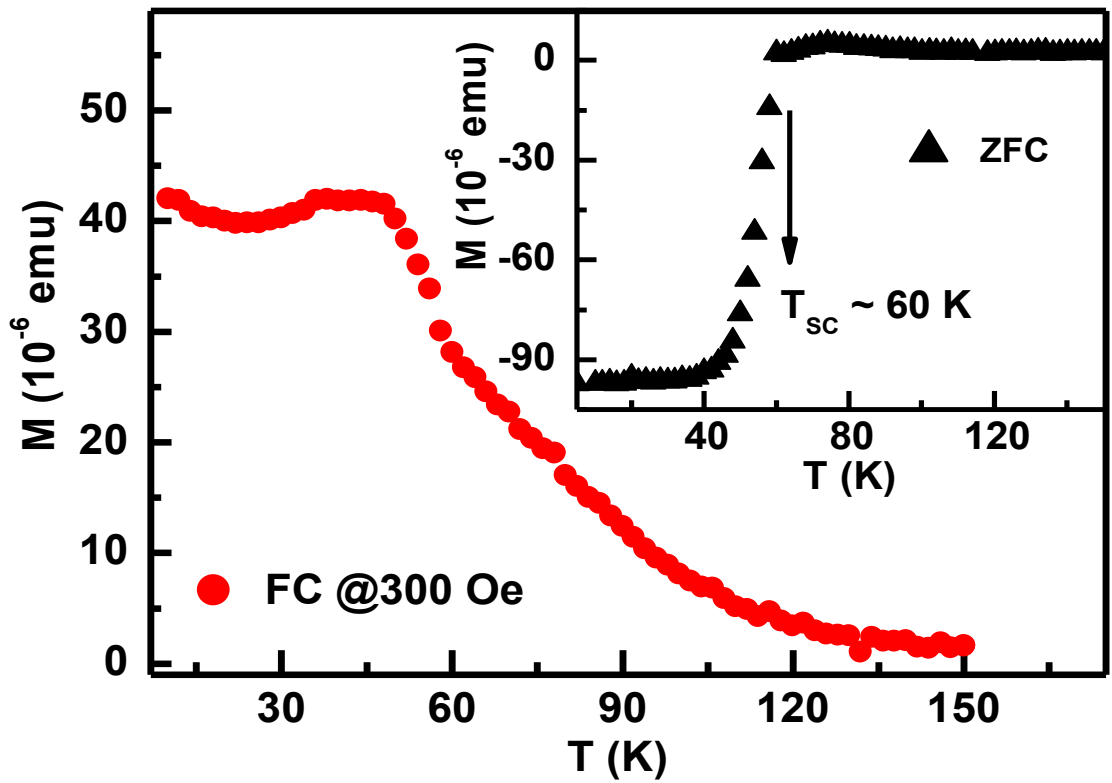

**Fig. 1**

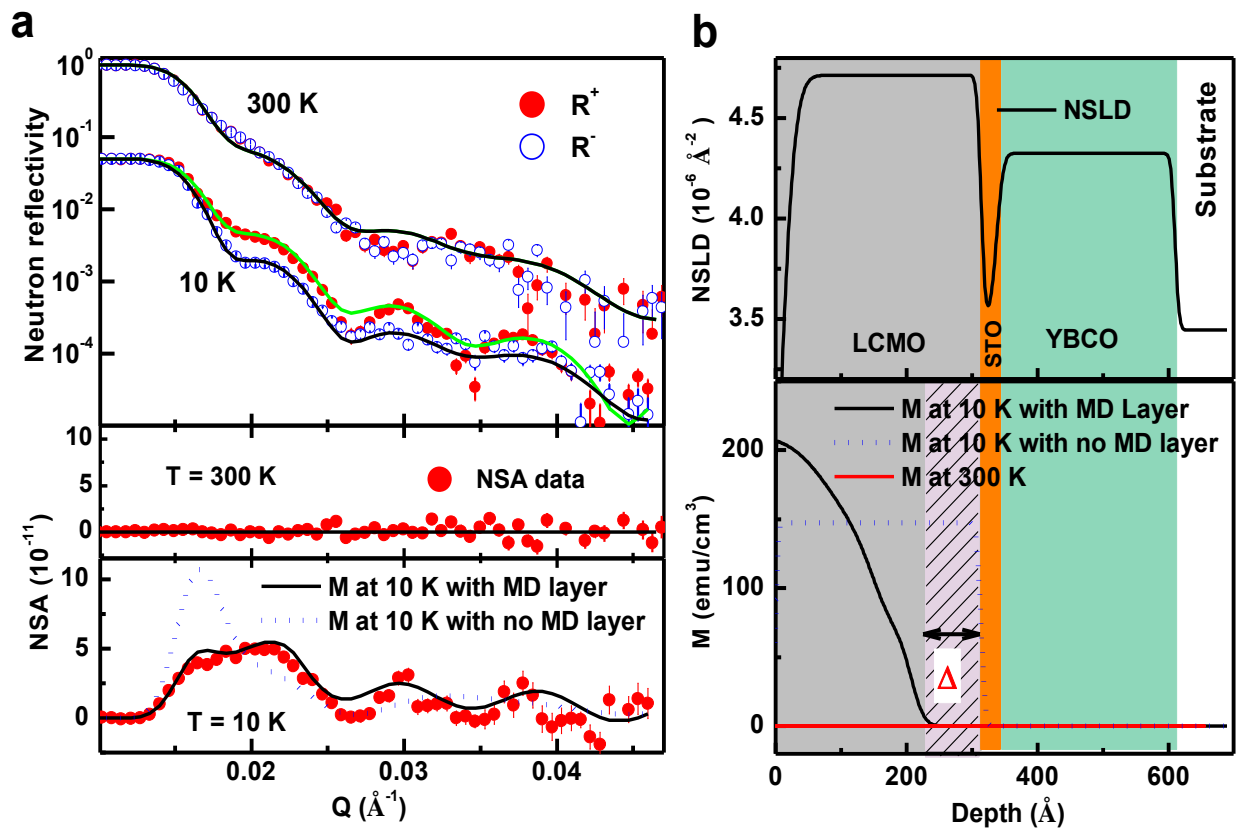

**Fig. 2**

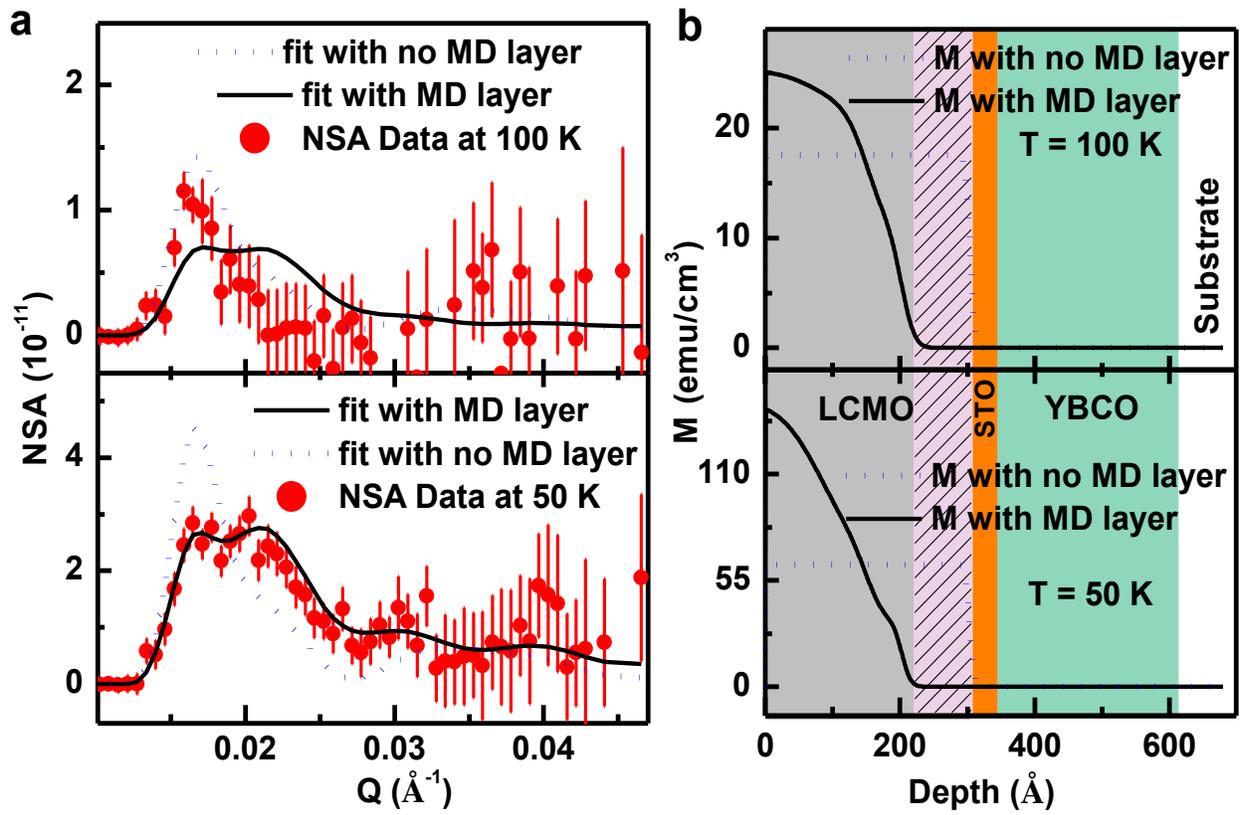

**Fig. 3**

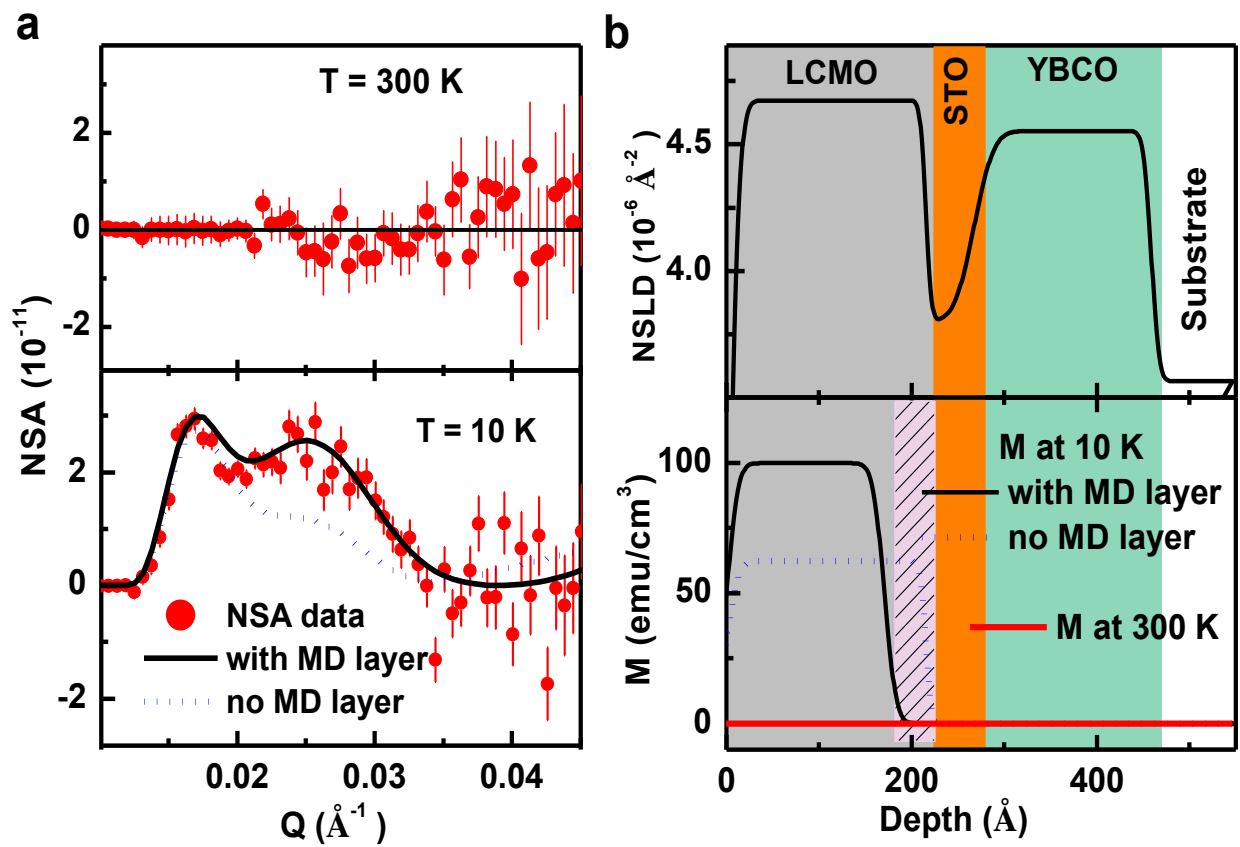

**Fig. 4**

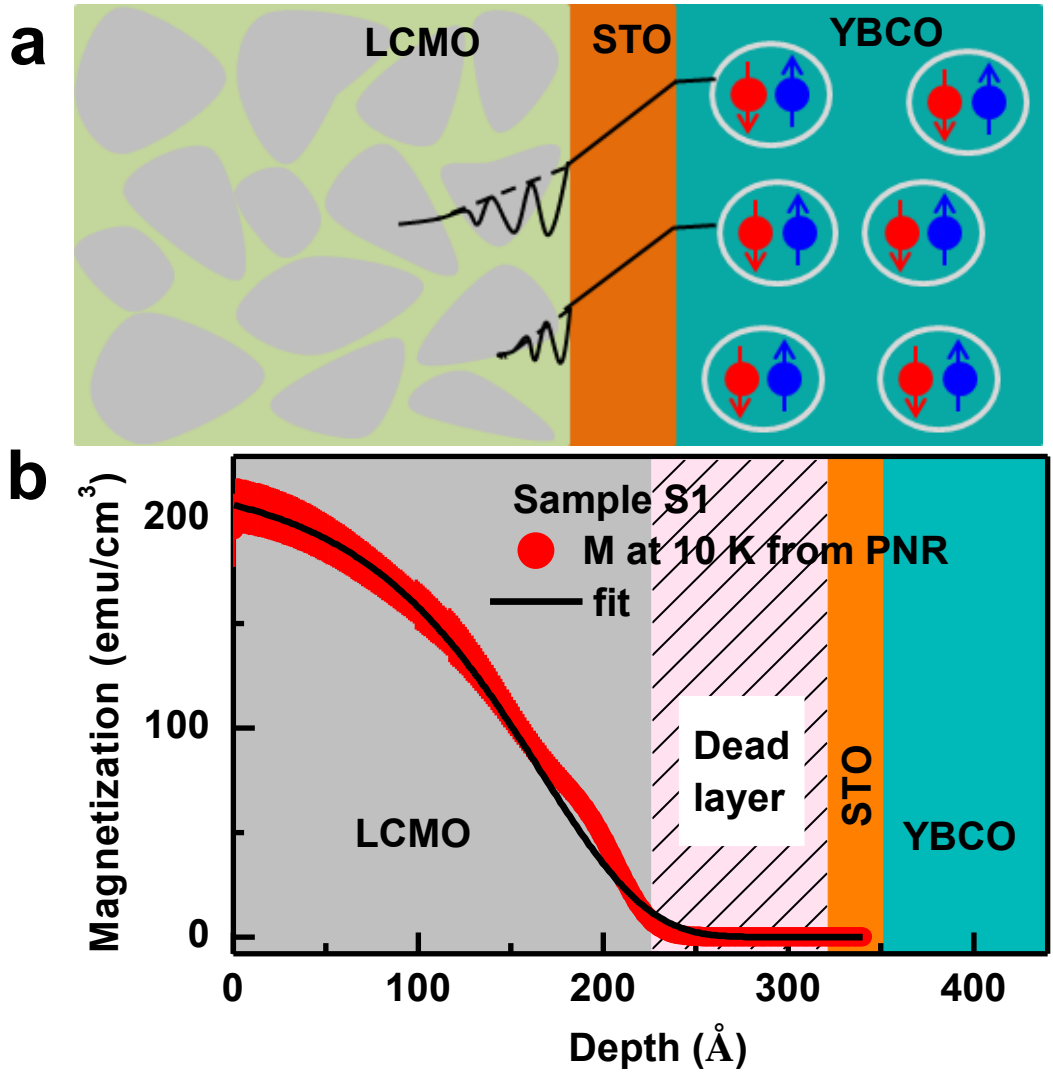

Fig 5

# Supplementary information for

# Superconductivity-induced Magnetic Modulation in a Ferromagnet Through an Insulator in $La_{2/3}Ca_{1/3}MnO_3/SrTiO_3/YBa_2Cu_3O_{7-\delta}$ Hybrid Heterostructures

C. L. Prajapat[1], Surendra Singh[2], Amitesh Paul[3], D. Bhattacharya[2], M. R. Singh[1], G. Ravikumar[1] and S. Basu[2,*]

[1]*Technical Physics Division, Bhabha Atomic Research Centre, Mumbai-400085.*

[2]*Solid State Physics Division, Bhabha Atomic Research Centre, Mumbai-400085.*

[3]*Technische Universitat Munchen, Physik Department E21, Lehrstuhl fur Neutronenstreuung, James-Franck-Strasse 1,D-85748 Garching b. München, Germany*

*email: sbasu@barc.gov.in

Pulsed laser (KrF) deposition was used to grow $La_{2/3}Ca_{1/3}MnO_3$ (LCMO)/ $SrTiO_3$ (STO)/ $YBa_2Cu_3O_{7-\delta}$ (YBCO) hetrostructure on single crystalline STO (001) substrates. The deposition rate was controlled through appropriate focus of laser beam on the target. The substrate temperature during film growth was initially optimized and was maintained at 770 °C. The oxygen pressure during deposition was 0.5 mbar. The laser fluence was 2.5 J/cm$^2$, and the pulse repetition rate was 2 Hz.

The degree of crystallinity of the hetrostructure was evaluated by X-ray Diffraction (XRD) measurements. Before deposition of hetrostructure we optimized the growth of LCMO and YBCO on single crystalline STO (001) substrate. Fig. S1 represents the XRD data from the sample. Upper and middle panels of the Fig. S1 show the XRD pattern from LCMO and YBCO layer grown on STO substrate. Lower panel of the Fig. S1 show the XRD pattern from the LCMO (300 Å)/ STO (25 Å)/YBCO (300 Å) hetrostructure sample. These results are evidence for a high degree of perfection of atomic structure along the growth direction.

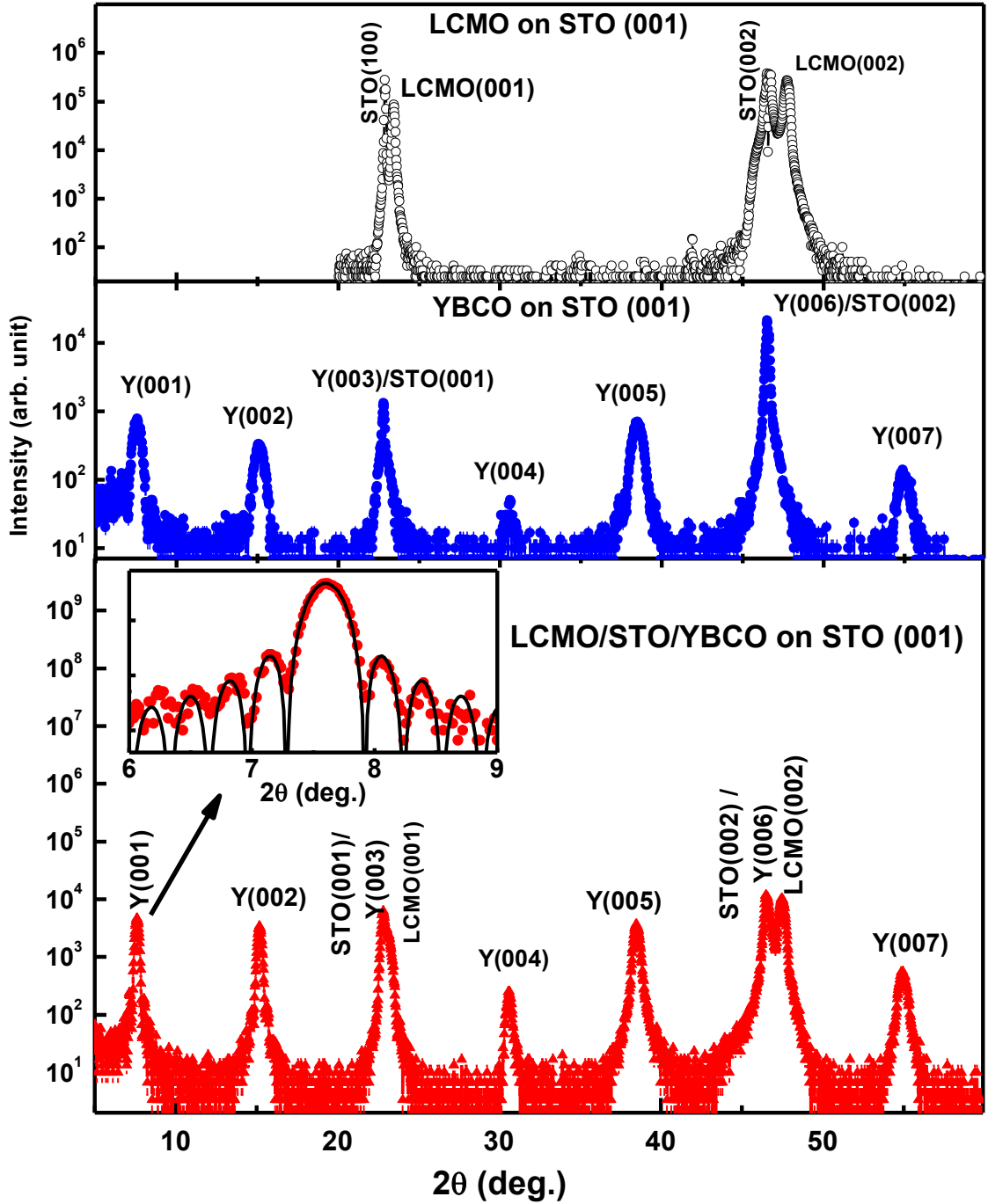

Fig. S1: X-ray diffraction data from LCMO/STO/YBCO system.

Fig. S2 shows the X-ray reflectivity (XRR) pattern from two hetrostructures, LCMO (300 Å)/STO (25 Å)/YBCO (300 Å) and LCMO (200 Å)/STO (50 Å)/YBCO (200 Å), samples. The specular reflectivity ($R$) was measured as a function of wave vector transfer, $Q = 4\pi \sin\theta/\lambda$ (where, $\theta$ is angle of incidence and $\lambda$ is x-ray). The reflectivity is qualitatively related to the Fourier transform of the scattering length density (SLD) depth profile $\rho(z)$[1,2], averaged over whole sample area. For XRR, $\rho(z)$, is proportional to electron density[1,2]. Thus the chemical depth profiles were inferred from the data by fitting a model $\rho(z)$ whose reflectivity best fit the data. The reflectivities were calculated using the dynamical formalism of Parratt[3], and parameters of the model were adjusted to minimize the value of reduced $\chi^2$ –a weighted measure of goodness of fit.[4] A model consisted of a layer(s) representing regions with different electron SLD. The parameters of the model included layer thickness, interface (or surface) roughness and electron SLD.

XRR pattern from LCMO (300 Å)/STO (25 Å)/YBCO (300 Å) and LCMO (200 Å)/STO (50 Å)/YBCO (200 Å) hetrostructure samples are shown in upper and lower panel of Fig. S2. Inset show the corresponding electron scattering length density (ESLD) profile which gave best fit to XRR data. The parameters obtained from the analysis of the XRR data are shown in Table 1.

Table 1 : parameters obtained from XRR measurements

|  | LCMO (300 Å)/STO (25 Å)/YBCO (300 Å) hetrostructure | LCMO (200 Å)/STO (50 Å)/YBCO (200 Å) hetrostructure |
|---|---|---|
|  |  |  |

| layer | Thickness (Å) | Electron SLD ($10^{-5}$ Å$^{-2}$) | Roughness (Å) | Thickness (Å) | Electron SLD ($10^{-5}$ Å$^{-2}$) | Roughness (Å) |
|---|---|---|---|---|---|---|
| LCMO | 320±15 | 4.97±0.06 | 13±3 | 187±12 | 4.88±0.07 | 10±3 |
| STO | 23±2 | 4.30±0.05 | 4±1 | 50±3 | 4.25±0.05 | 5±1 |
| YBCO | 285±15 | 4.76±0.04 | 12±4 | 185±11 | 4.74±0.05 | 14±4 |

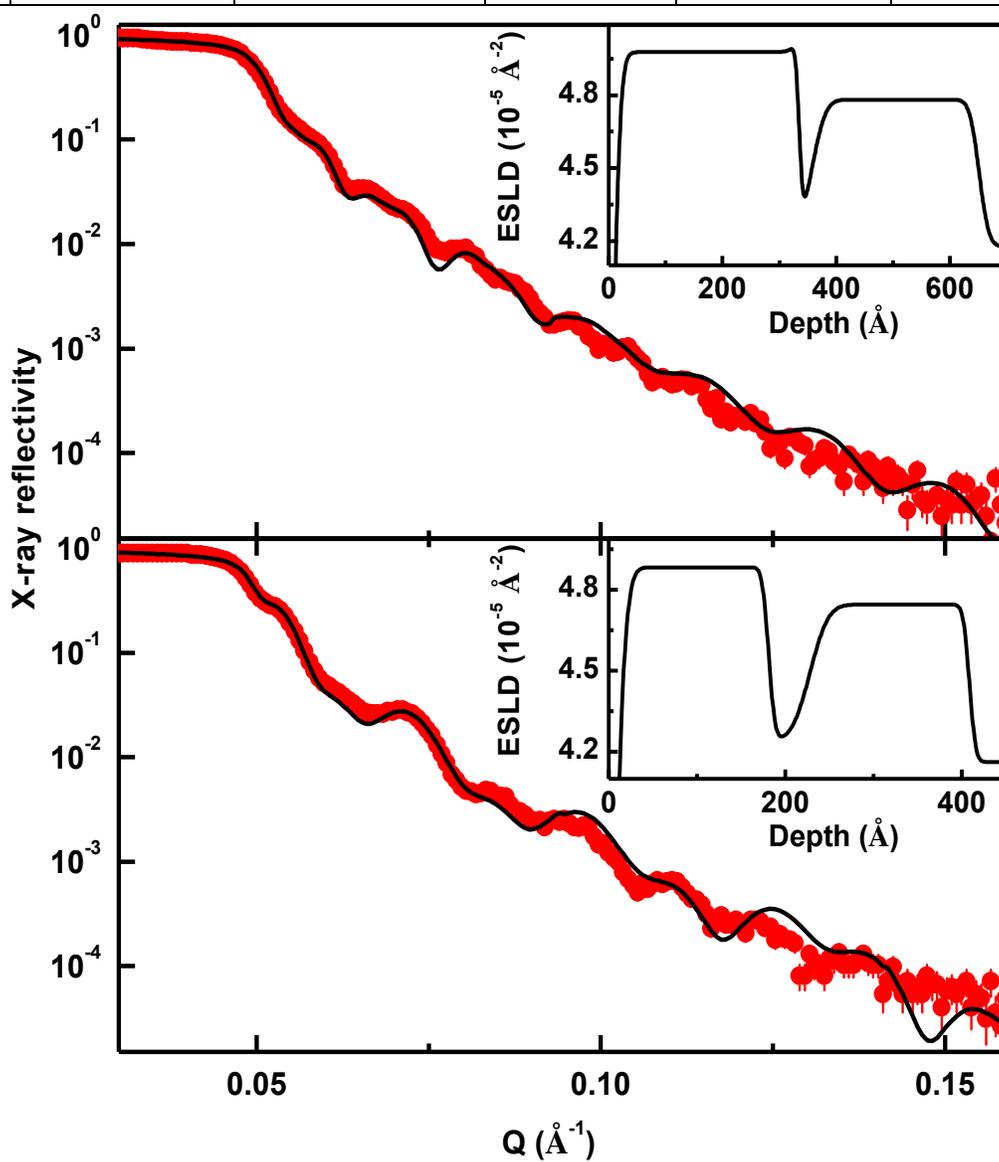

Fig. S2: X-ray reflectivity (XRR) pattern from LCMO/STO/YBCO hetrostructures. Inset show the corresponding electron scattering length density (ESLD) depth profile which gave best fit to XRR data.

Fig. S3 show the polarized neutron reflectivity (PNR) data from LCMO (300 Å)/STO (25 Å)/YBCO (300 Å) hetrostructure at 10 K. We first optimized the nuclear scattering length density (or NSLD) profile from PNR data at 300 K (where there is no magnetism) by constraining layer thicknesses and interface roughness to be within the 95% confidence limit,[4] i.e., 2-$\sigma$ error, established from the analysis of the XRR data. To fit PNR data at 10 K we optimized magnetization depth profile only and NSLD profile was fixed. Fig. S3 a show the PNR data at 10 K. upper panels show the spin difference ($R^+ - R^-$) data.

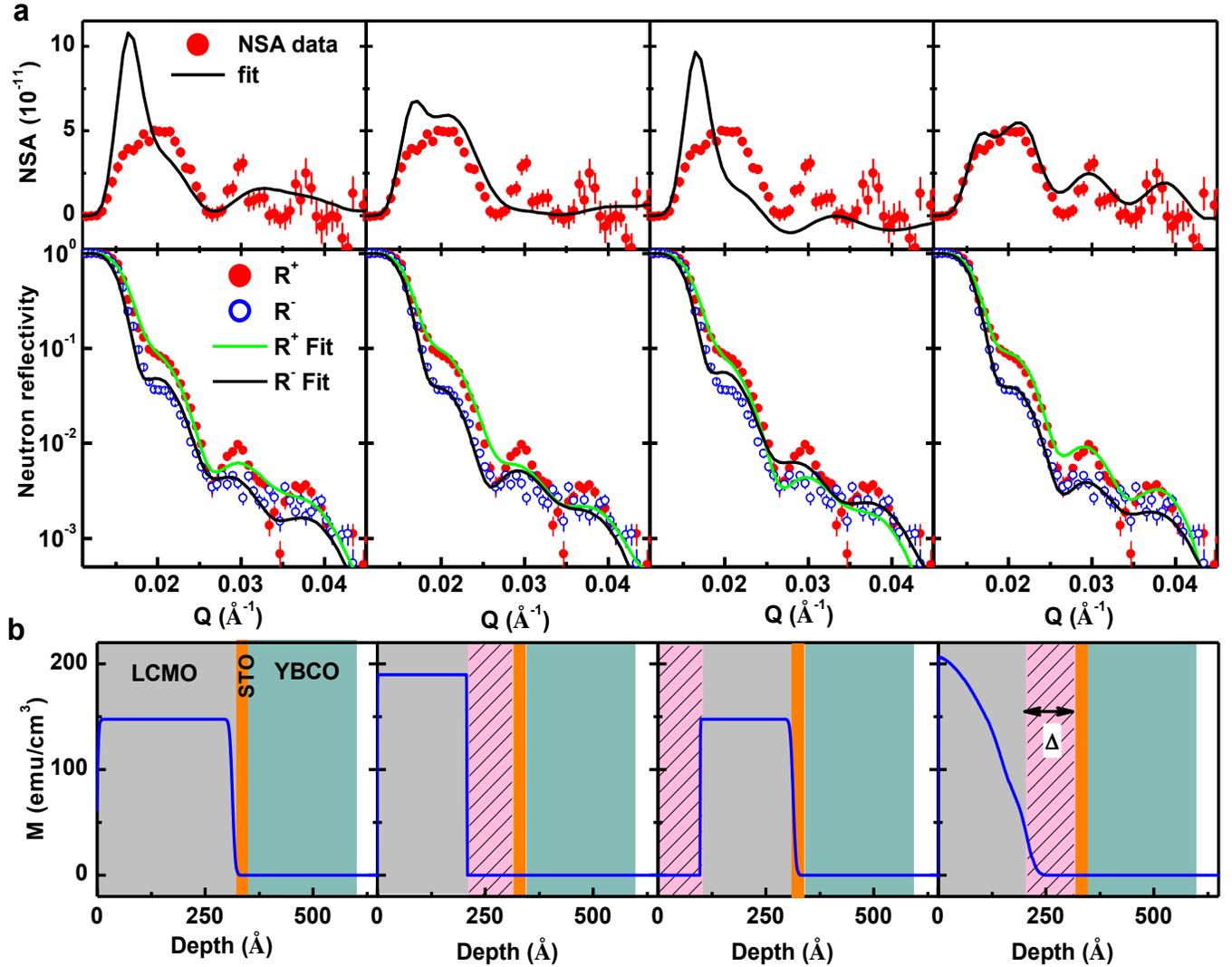

Fig. S3: **a,** PNR (spin up, $R^+$ and spin down, $R^-$) data from the YBa$_2$Cu$_3$O$_{7-\delta}$ (300Å)/SrTiO$_3$ (25Å)/ La$_{2/3}$Ca$_{1/3}$MnO$_3$ (300 Å) sample at 10 K. upper panel show the normalized spin asymmetry (NSA) [ =(R$^+$ - R$^-$)/R$_F$, where R$_F$ is Fresnel reflectivity] data at 10 K. **b,** shows the corresponding magnetization (*M*) depth profiles which fitted PNR data at 10 K.

We used different models for the magnetization depth profile by considering uniform and non uniform magnetization across LCMO layer. A comparison of three models which gave better fit (with smaller $\chi^2$) to PNR data at 10 K are shown in Fig. S3b. These three models are (a)

Where the magnetization is homogeneous throughout LCMO layer (left panel), (b) magnetization is suppressed (or formation of magnetic dead layer) at LCMO/STO interface but uniform magnetization in the rest of LCMO layer (middle panel) and (c) formation of magnetic dead layer at LCMO/STO interface and non uniform magnetization in the rest of LCMO layer (right panel). Fig S3 clearly depicts that model (c) best fit (with smallest $\chi^2$) the PNR data at 10 K, suggesting modulation in magnetization depth profile LCMO layer.

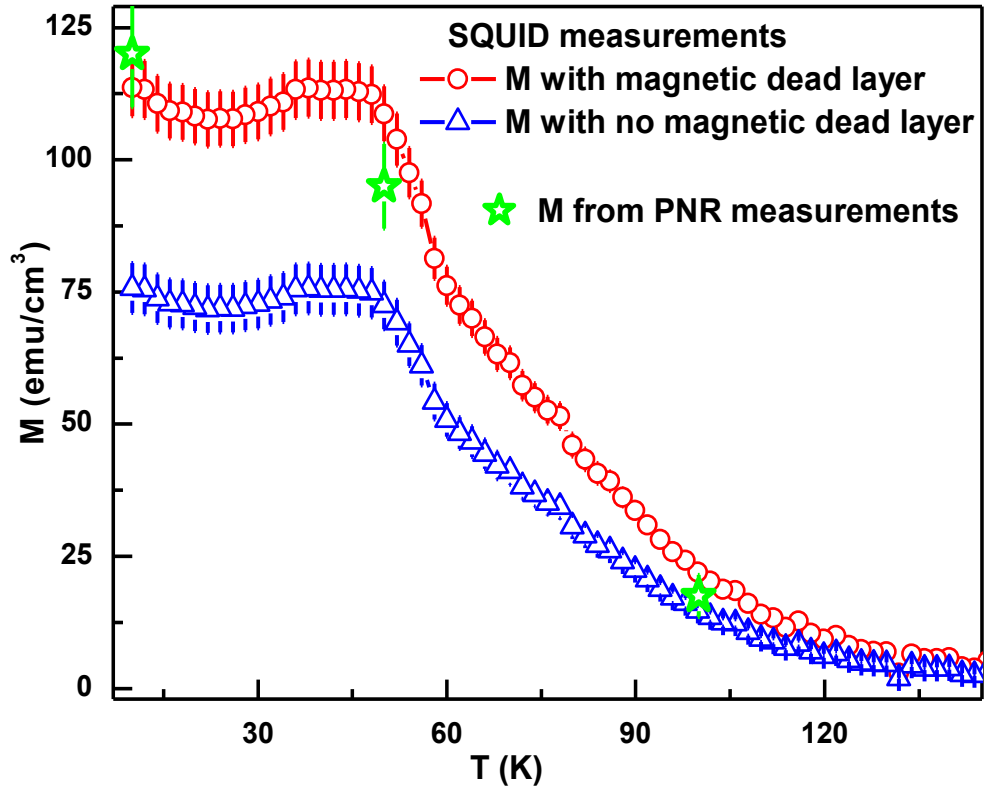

Fig. S4: Variation of magnetization (M) as a function of temperature for field cooled condition in a magnetic field of 300 Oe.

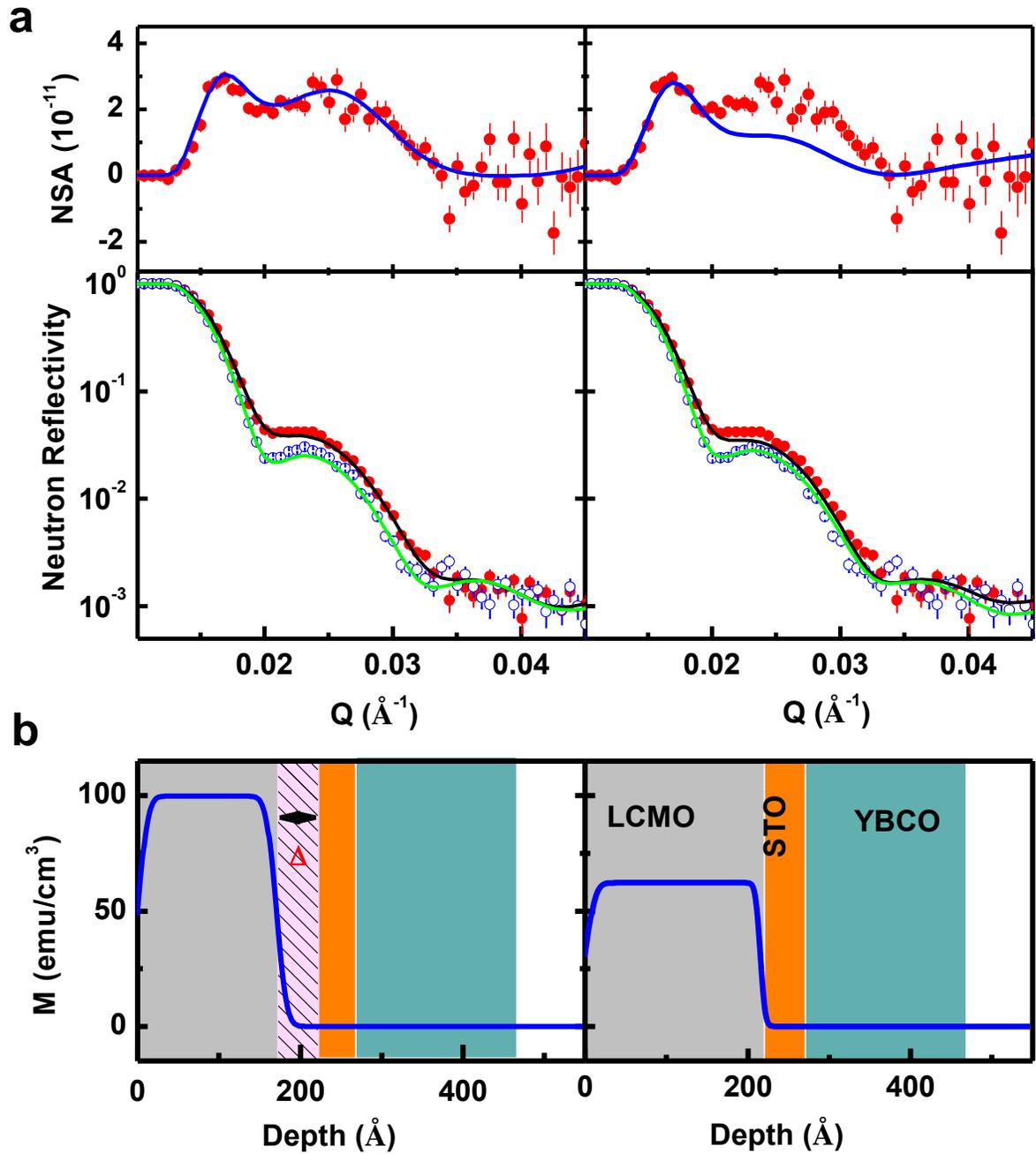

Fig. S5: **a,** PNR (spin up, $R^+$ and spin down, $R^-$) data from the YBa$_2$Cu$_3$O$_{7-\delta}$ (200 Å)/SrTiO$_3$ (50 Å)/La$_{2/3}$Ca$_{1/3}$MnO$_3$ (200 Å) sample at 10 K. upper panel show the normalized spin asymmetry (NSA) [= $(R^+ - R^-)/R_F$, where $R_F$ is Fresnel reflectivity] data at 10 K. **b,** shows the corresponding magnetization depth profiles which fitted PNR data at 10 K.

The variation of magnetization (M in emu/cm$^3$) as a function of temperature is determined from SQUID measurements are shown in Fig. S4. Using thickness of LCMO layer as measured by scattering techniques (XRR and PNR) we obtained magnetization in emu/cc and are plotted in Fig. S4 assuming two cases: with (open circle with line) and without (open triangle with line) magnetic dead layer in LCMO layer. The magnetization obtained from SQUID on assuming a magnetic dead layer of thickness ~ 100 Å matches well with the ones obtained from the best fit of PNR data (shown by open star).

Fig. S5 show PNR data analysis assuming different magnetization profile for sample S2 at 10 K. It is clear from the Fig. S5 that magnetic dead layer model best fit the PNR data.